\documentstyle[12pt]{article}
\newcommand{\Frac}[2]%
{{\textstyle \frac{\mbox{\footnotesize $#1$}\rule[-0.9mm]{0mm}{1mm}}%
{\mbox{\footnotesize $#2$}\rule{0mm}{3.1mm}}}}
\begin{document}
\begin{titlepage}
\vspace*{-12 mm}
\noindent
\begin{flushright}
\begin{tabular}{l@{}}
hep-ph/9712507 \\
\end{tabular}
\end{flushright}
\vskip 12 mm
\begin{center}
{\Large \bf A polarised QCD condensate:} \\
\vspace{3ex}
{\Large $\nu p$ elastic scattering as a probe of $U_A(1)$ dynamics} 
\\[14 mm]
{\bf S.D. Bass}\footnote{Steven.Bass@mpi-hd.mpg.de}
\\[10mm]   
{\em Institut f\"{u}r Theoretische Kernphysik, 
Universit\"{a}t Bonn,\\
Nussallee 14--16, D-53115 Bonn, Germany}\\[5mm]
and \\[5mm]
{\em \footnote{Present address} 
Max Planck Institut f\"ur Kernphysik,\\
Postfach 103980, 
D-69029 Heidelberg, Germany}

\end{center}
\vskip 10 mm
\begin{abstract}
\noindent
$U_A(1)$ dynamics have the potential to induce a polarised
condensate inside a nucleon.
The formation of this condensate is related 
to the realisation of 
$U_A(1)$ symmetry breaking by tunneling processes such as
instantons.
If it is present, the polarised condensate induces a term
in $g_1$
which has support only at $x=0$.
Tunneling processes then induce a net transfer of ``spin'' 
from finite $x$ to $x=0$.
The polarised condensate may be measured by comparing 
the flavour-singlet axial charges which are extracted 
from polarised deep inelastic and $\nu p$ elastic scattering 
experiments.
\end{abstract}
\end{titlepage}
\renewcommand{\labelenumi}{(\alph{enumi})}
\renewcommand{\labelenumii}{(\roman{enumii})}
\newpage
%

\section{Introduction}

Polarised deep inelastic scattering experiments at CERN
\cite{expt,smc},
DESY \cite{hermes} and SLAC \cite{e143a,e154}
have revealed an apparent
four standard deviations violation of OZI 
in the flavour-singlet axial charge $g_A^{(0)}$
which is extracted from the first moment of $g_1$ 
(the nucleon's first spin dependent structure function).
This result is the 
EMC Spin Effect.
It has inspired many QCD based explanations involving
the parton model \cite{efremov,ccm} and $U_A(1)$ dynamics
\cite{venez,forte2,ioffe}.

The topological structure of the QCD vacuum 
\cite{Callan,rjc} is believed to play an important 
role in the physics of the $U_A(1)$
channel \cite{rjc}
through the axial anomaly \cite{adler,bell}.
In this paper 
we explain how tunneling processes between vacuum states
with different topological winding number
may generate a polarised condensate {\it inside a nucleon}.
Whether this condensate forms or not
is related to the realisation of $U_A(1)$ symmetry breaking 
\cite{rjc,thooft,thrept}
by
tunneling processes such as instantons.
If a polarised condensate does form,
then the vacuum
inside a nucleon
acquires a net axial charge relative to the vacuum outside a nucleon.
The condensate induces a term in $g_1$
which has support only at Bjorken $x$ equal to zero.

Polarised deep inelastic scattering experiments measure
$g_1(x,Q^2)$ between some small but finite value $x_{\rm min}$
and an upper value $x_{\rm max}$ which is close to one.
As we decrease $x_{\rm min} \rightarrow 0$
we measure the first moment
\begin{equation}
\Gamma \equiv \lim_{x_{\rm min} \rightarrow 0} \ 
\int^1_{x_{\rm min}} dx \ g_1 (x,Q^2).
\end{equation}
Polarised deep inelastic experiments
cannot, even in principle, measure at $x=0$ with finite $Q^2$.
As noted in \cite{AL},
they miss any 
$\delta (x)$ terms which might exist in $g_1$ at large $Q^2$.
Suppose that a polarised condensate exists and that it
contributes an amount
$\lambda$ to the flavour-singlet axial charge $g_A^{(0)}$.
The flavour-singlet axial charge
which is 
extracted from a polarised deep inelastic experiment is
$(g_A^{(0)} - \lambda)$.
In contrast, elastic ${\rm Z}^0$ exchange processes such as
$\nu p$ elastic scattering \cite{garvey}
measure the full $g_A^{(0)}$ \cite{bcft,kaplan}.
One can measure a polarised condensate
by comparing the flavour-singlet axial charges
which are extracted from polarised deep inelastic and
$\nu p$ elastic scattering experiments.

The structure of the paper is as follows.
We first review (Section 2)
the role of the axial anomaly in QCD based explanations of 
the first moment of $g_1$.
We distinguish two infra-red problems:
the factorisation scheme dependence of the QCD parton model
\cite{bint}
(a problem in perturbative QCD)
and the need to ensure that the theory 
is invariant under
topologically non-trivial gauge transformations \cite{jaffe,forte1,bass}.
The latter requirement leads us to consider the possibility 
that a
polarised condensate may form inside a nucleon.
In Section 3 we outline a dynamical mechanism 
for the formation
of this condensate.
We compare the two mechanisms of spontaneous \cite{rjc}
and explicit \cite{thooft}
breaking of $U_A(1)$ symmetry by instantons.
We find that spontaneous symmetry breaking induces a
polarised condensate whereas explicit symmetry breaking
does not.
Finally (Section 4), we discuss experiments which might be used
to look for a polarised condensate in QCD.

We note that polarised condensates are observed
in low temperature physics.
The A phase of superfluid $^3$He behaves both as an orbital
ferromagnet and uniaxial liquid crystal with spontaneous 
magnetisation along the anistropy axis ${\hat l}$,
and as a spin antiferromagnet with magnetic anisotropy along a second
axis ${\hat d}$ \cite{anderson}.

\section {The axial anomaly and $g_1$}

\subsection{The first moment of $g_1$}

When $Q^2 \rightarrow \infty$,
the light-cone operator product expansion relates
the first moment of the structure function $g_1$
to the scale-invariant axial charges of the target nucleon
by \cite{kod,larin,altr}
\begin{eqnarray}
\int_0^1 dx \ g_1^p (x,Q^2) &=&
\Biggl( {1 \over 12} g_A^{(3)} + {1 \over 36} g_A^{(8)} \Biggr)
\Bigl\{1 + \sum_{\ell\geq 1} c_{{\rm NS} \ell\,}
\bar{g}^{2\ell}(Q)\Bigr\} \nonumber \\
&+& {1 \over 9} g_A^{(0)}|_{\rm inv}
\Bigl\{1 + \sum_{\ell\geq 1} c_{{\rm S} \ell\,}
\bar{g}^{2\ell}(Q)\Bigr\}
\ + \ {\cal O}({1 \over Q^2}).
\end{eqnarray}
Here $g_A^{(3)}$, $g_A^{(8)}$ and $g_A^{(0)}|_{\rm inv}$ are the 
isotriplet, SU(3) octet and scale-invariant 
flavour-singlet axial charges respectively.
The flavour non-singlet $c_{{\rm NS} \ell}$
and singlet $c_{{\rm S} \ell}$ coefficients
are calculable in
$\ell$-loop perturbation theory \cite{larin}.

The first moment of $g_1$ is fully constrained
by low energy
weak interaction dynamics.
For proton states $|p,s\rangle$ with momentum $p_\mu$ and spin $s_\mu$
\begin{eqnarray}
2 m s_{\mu} \ g_A^{(3)} &=&
\langle p,s | 
\left(\bar{u}\gamma_\mu\gamma_5u - \bar{d}\gamma_\mu\gamma_5d \right)
| p,s \rangle _c  \nonumber \\
2 m s_{\mu} \ g_A^{(8)} &=&
\langle p,s |
\left(\bar{u}\gamma_\mu\gamma_5u + \bar{d}\gamma_\mu\gamma_5d
                   - 2 \bar{s}\gamma_\mu\gamma_5s\right)
| p,s \rangle _c
\end{eqnarray}
where the subscript $c$ denotes the connected matrix element.
The isotriplet axial charge $g_A^{(3)}$ is measured independently
in neutron beta decays and, modulo ${\rm SU_F}(3)$ breaking
\cite{close},
the flavour octet axial charge
$g_A^{(8)}$ is measured independently in hyperon beta decays.
The scale-invariant flavour-singlet axial charge 
$g_A^{(0)}|_{\rm inv}$ 
is defined by \cite{mink}
\begin{equation}
2m s_\mu g_A^{(0)}|_{\rm inv} = 
\langle p,s|
E(g)J^{GI}_{\mu5}(z) |p,s\rangle _c
\label{e}\end{equation}
where 
\begin{equation}
J^{GI}_{\mu5} = \left(\bar{u}\gamma_\mu\gamma_5u
                  + \bar{d}\gamma_\mu\gamma_5d
                  + \bar{s}\gamma_\mu\gamma_5s\right)_{GI}
\label{aa}\end{equation} 
is the
gauge-invariantly renormalized singlet axial-vector operator 
and
\begin{equation}
E(g) = \exp \int^g_0 \! dg'\, \gamma(g')/\beta(g')
\label{a}\end{equation}
is a renormalisation group factor;
$\beta(g)$ and $\gamma(g)$ are the 
Callan-Symanzik functions associated with the 
gluon coupling constant $g$ and the composite operator 
$J^{GI}_{\mu5}$ respectively.
We are free to choose the coupling $g(\mu)$ at either a 
hard or a soft scale $\mu$.
The singlet axial charge 
$g_A^{(0)}|_{\rm inv}$ 
is independent of the renormalisation scale $\mu$.
It may be measured independently in an elastic 
neutrino proton scattering experiment \cite{bcft,kaplan}.

Polarised deep inelastic scattering experiments
at CERN \cite{expt,smc}, DESY \cite{hermes} and SLAC
\cite{e143a,e154}
have verified
the Bjorken sum-rule \cite{bj}
for the isovector part of $g_1$
to within 15\%.
They have also revealed an apparant four standard deviations 
violation 
of OZI in the flavour singlet axial charge $g_A^{(0)}|_{\rm inv}$
---
for recent reviews see \cite{altr,cheng}.
The current experimental value of 
\begin{equation}
g_A^{(0)}|_{\rm inv}
= \Delta u_{\rm inv} + \Delta d_{\rm inv} + \Delta s_{\rm inv}
\end{equation}
from polarised deep inelastic scattering
is \cite{badelek}
\begin{equation}
\left. g^{(0)}_A \right|_{\rm inv} = 0.28 \pm 0.07. 
\label{fa}\end{equation}
This value is extracted assuming no $\delta (x)$ term in $g_1$.
It compares with $g_A^{(8)} = 0.58 \pm 0.03$ 
from hyperon beta-decays \cite{close}.
Deep inelastic measurements of
$g_A^{(0)}|_{\rm inv}$ 
involve a
smooth extrapolation of the $g_1$ data to $x=0$ 
which is motivated either by
Regge theory or by perturbative QCD.
The
small $x$ extrapolation of $g_1$ data
is presently the largest source of experimental error on
measurements of the nucleon's axial charges from deep inelastic
scattering.

The OZI violation in $g_A^{(0)}|_{\rm inv}$ has a natural 
interpretation in terms
of the axial anomaly \cite{adler,bell} in QCD.
We now briefly review the theory of the axial anomaly in
nucleon matrix elements of 
$J^{GI}_{\mu5}$ (Section 2.2) and
and explain its application to the nucleon's internal spin structure
(Section 2.3).

\subsection{The axial anomaly}

The gauge invariant current $J_{\mu 5}^{GI}$ satisfies the
anomalous divergence equation
\begin{equation}
\partial^\mu J^{GI}_{\mu5}
= 2f\partial^\mu K_\mu + \sum_{i=1}^{f} 2im_i \bar{q}_i\gamma_5 q_i
\end{equation}
where
\begin{equation}
K_{\mu} = {g^2 \over 16 \pi^2}
\epsilon_{\mu \nu \rho \sigma}
\biggl[ A^{\nu}_a \biggl( \partial^{\rho} A^{\sigma}_a 
- {1 \over 3} g 
f_{abc} A^{\rho}_b A^{\sigma}_c \biggr) \biggr]
\end{equation}
is a renormalized version of the gluonic Chern-Simons
current
and the number of light flavours $f$ is $3$.

Equ.(9) allows us to write
\begin{equation}
J_{\mu 5}^{GI} = J_{\mu 5}^{\rm con} + 2f K_{\mu}
\end{equation}
where $J_{\mu 5}^{\rm con}$ and $K_{\mu}$ satisfy the
divergence equations
\begin{equation}
\partial^\mu J^{\rm con}_{\mu5}
= \sum_{i=1}^{f} 2im_i \bar{q}_i\gamma_5 q_i
\end{equation}
and
\begin{equation}
\partial^{\mu} K_{\mu} 
= {g^2 \over 8 \pi^2} G_{\mu \nu} {\tilde G}^{\mu \nu}.
\end{equation}
Here
${g^2 \over 8 \pi^2} G_{\mu \nu} {\tilde G}^{\mu \nu}$
is the topological charge density.
$J_{\mu 5}^{GI}$ is the only 
flavour-singlet,
gauge-invariant,
twist-two operator with $J^{PC}=1^{++}$. 
It is multiplicatively renormalised with a 
finite anomalous
dimension $\gamma (g)$ which starts at two loops in perturbation
theory.
The partially conserved current is scale invariant, viz.
\begin{equation}
J_{\mu 5}^{\rm con} |_{\mu^2} = J_{\mu 5}^{\rm con} |_{\mu_0^2}.
\end{equation}
It follows that the scale dependence of $J_{\mu 5}^{GI}$ is
carried entirely by $K_{\mu}$.
One finds
\begin{equation}
K_{\mu}|_{\mu^2} = Z(\mu^2, \mu_0^2) K_{\mu}|_{\mu_0^2} +
                   \biggl( Z(\mu^2, \mu_0^2) -1 \biggr) J_{\mu 5}^{\rm con} 
\end{equation}
where
\begin{equation}
Z(\mu^2, \mu_0^2)
= \exp \int_{g(\mu_0^2)}^{g(\mu^2)} dg' \gamma (g') / \beta (g') 
\end{equation}
is the renormalisation group factor;
$Z(\mu^2, \mu_0^2) \rightarrow E(g)$ in the limit that 
$\mu_0 \rightarrow \infty$.

When we make a gauge transformation $U$ 
the gluon field transforms as
\begin{equation}
A_{\mu} \rightarrow U A_{\mu} U^{-1} + {i \over g} (\partial_{\mu} U) U^{-1}
\end{equation}
and the operator $K_{\mu}$
transforms as
\begin{equation}
K_{\mu} \rightarrow K_{\mu} 
+ i {g \over 16 \pi^2} \epsilon_{\mu \nu \alpha \beta}
\partial^{\nu} 
\biggl( U^{\dagger} \partial^{\alpha} U A^{\beta} \biggr)
+ {1 \over 96 \pi^2} \epsilon_{\mu \nu \alpha \beta}
\biggl[ 
(U^{\dagger} \partial^{\nu} U) 
(U^{\dagger} \partial^{\alpha} U)
(U^{\dagger} \partial^{\beta} U) 
\biggr].
\end{equation}
Gauge transformations shuffle a scale invariant operator quantity
between the two operators $J_{\mu 5}^{\rm con}$ and $K_{\mu}$
whilst keeping $J_{\mu 5}^{GI}$ invariant.
Since $K_{\mu}$ satisfies the divergence equation (13) it follows
that the $U$ dependent terms on the right hand side of (18) must
be conserved.
The second term on the right hand side of (18) is clearly conserved
by itself, whence we conclude that
\begin{equation}
{\cal C}_{\mu} =
 {1 \over 96 \pi^2} \epsilon_{\mu \nu \alpha \beta}
\biggl[ 
(U^{\dagger} \partial^{\nu} U) 
(U^{\dagger} \partial^{\alpha} U)
(U^{\dagger} \partial^{\beta} U) 
\biggr]
\end{equation}
is also conserved.

The nucleon matrix element of $J_{\mu 5}^{GI}$ is
\begin{equation}
\langle p,s|J^{GI}_{5 \mu}(0)|p',s\rangle 
= 2m \biggl[ s_\mu G_A (l^2) + l_\mu l.s G_P (l^2) \biggr]
\end{equation}
where $l_{\mu} = (p'-p)_{\mu}$.
Since $J^{GI}_{5 \mu}(0)$ does not couple to a massless 
Goldstone
boson it follows that $G_A(l^2)$ and $G_P(l^2)$ contain
no massless pole terms.
The forward matrix element of $J^{GI}_{5 \mu}(0)$ is well
defined and
\begin{equation}
g_A^{(0)}|_{\rm inv} = E(g) G_A (0).
\end{equation}

The matrix elements of $K_{\mu}$ need to be specified with
respect to a specific gauge.
In a covariant gauge we can write
\begin{equation}
\langle p,s|K_\mu(0)|p',s\rangle _c
= 2m \biggl[ s_\mu K_A(l^2) + l_\mu l.s K_P(l^2) \biggr]
\end{equation}
where $K_P$ contains a massless Kogut-Susskind pole \cite{kogut}.
This massless pole cancels with a corresponding massless
pole term in $(G_P - K_P)$.
In an axial gauge $n.A=0$ the matrix elements of the gauge dependent 
operator $K_{\mu}$ will also contain terms proportional to the gauge
fixing vector $n_{\mu}$.

We may define a gauge-invariant form-factor $\chi^{g}(l^2)$
for the topological charge density (13) in the divergence of 
$K_{\mu}$:
\begin{equation}
2m l.s \chi^g(l^2) =
\langle p,s | {g^2 \over 8 \pi^2} G_{\mu \nu} {\tilde G}^{\mu \nu}
(0) | p', s \rangle_c.
\end{equation}
Working in a covariant gauge, we find
\begin{equation}
\chi^{g}(l^2) = K_A(l^2) + l^2 K_P(l^2)
\end{equation}
by contracting Eq.(22) with $l^{\mu}$.
When we make a gauge transformation any change 
$\delta_{\rm gt}$
in $K_A(0)$ is compensated
by a corresponding change in the residue of the Kogut-Susskind
pole in $K_P$, viz.
\begin{equation}
\delta_{\rm gt} [ K_A(0) ]
+ \lim_{l^2 \rightarrow 0} \delta_{\rm gt} [ l^2 K_P(l^2) ] = 0.
\end{equation}
The residue of the Kogut-Susskind pole and, hence, $K_A(0)$
is invariant under the ``small'' gauge transformations of
perturbative QCD
\footnote{The Kogut-Susskind pole corresponds to the Goldstone
 boson associated with spontaneously broken $U_A(1)$ symmetry.
 There is no Kogut-Susskind pole in perturbative QCD.}.
(``Small'' gauge transformations are those which are topologically 
 deformable to the identity.)
It changes only under ``large'' gauge transformations which change
the gluonic boundary conditions at infinity or, equivalently, 
change the topological winding number.

The functional variation of the integral of ${\cal C}_{\mu}[U]$ 
over a three dimensional submanifold ${\cal V}$ 
in Minkowski space can be written as an integral 
over the boundary
$\partial {\cal V}$ of ${\cal V}$ \cite{cron}.
The integral is insensitive to local deformations of $U$.
In contrast,
the second term on the right hand side of (18) and its integral
are sensitive to local deformations of both $A_{\mu}$ and $U$. 
This second term is a total divergence and 
its matrix elements vanish in the forward direction.
The matrix elements of ${\cal C}_{\mu}$ 
are sensitive to topological structure. 
They include the Kogut-Susskind pole.
Any change in $K_A(0)$ 
under a ``large'' gauge transformation is associated 
with ${\cal C}_{\mu}$.

To summarise, in a covariant gauge the forward matrix elements
of $K_{\mu}$ change under ``large'' gauge transformations but
not under ``small'' gauge transformations.
This change is associated with ${\cal C}_{\mu}$, which acts as
a topological current for gauge transformations \cite{cron}.

One can find axial gauges where the forward matrix element of
a given component of $K_{\mu}$ is invariant under residual
gauge transformations.
In the light-cone gauge $A_+=0$ the non-abelian three-gluon 
part of $K_+$ and, hence, ${\cal C}_+$ vanishes.
The forward matrix elements of $K_+$ are invariant under all
residual gauge degrees of freedom in the light-cone gauge.

\subsection{The anomaly and the first moment of $g_1$}

We now discuss the application of this theory to polarised
deep inelastic scattering.
The structure function $g_1$ appears in the anti-symmetric,
spin dependent part of the hadronic tensor $W^{\mu \nu}$
for $ep$ scattering,
viz.
\begin{equation}
{1 \over 2m} W^{\mu \nu}_A =
i \epsilon^{\mu \nu \rho \sigma} q_{\rho} s_{\sigma}
\biggl(
{1 \over p.q} g_1 (x,Q^2)
+ [ p.q s_{\sigma} - s.q p_{\sigma} ] {1 \over m^2 p.q} g_2 (x,Q^2) \biggr).
\end{equation}
Let $q_{\mu}$ denote the momentum of the exchanged photon.
The kinematics of polarised deep inelastic scattering are
$Q^2 = - q^2$ and $p.q$ both $\rightarrow \infty$ with 
$x = {Q^2 \over 2 p.q}$ held fixed.
In light-cone coordinates $q_- \rightarrow \infty$ with
$q_+ = -x p_+$ finite.
In this Bjorken limit the leading term in $W_A^{\mu \nu}$
is obtained by taking $\sigma = +$.
(Other terms are suppressed by powers of ${1 \over q_-}$.)
Understanding the first moment of $g_1$ in terms of the matrix
elements of anomalous currents
($J_{\mu 5}^{\rm con}$ and $K_{\mu}$)
is a problem in understanding the forward matrix element of $K_+$.

Here we are fortunate in that the parton model is formulated in
the light-cone gauge where the forward matrix elements of $K_+$
are invariant.
Furthermore, in this gauge, $K_+$ measures the gluonic ``spin'' 
content
of the polarised target \cite{jafpl,man90}.
We find \cite{efremov,ccm}
\begin{equation}
G_A^{(\rm A_+ = 0)}(0) = \sum_q \Delta q_{\rm con} 
- f {\alpha_s \over 2 \pi} \Delta g
\end{equation}
where
\begin{equation}
\langle p,s|J_{+5}^{\rm con}(0)|p,s\rangle _c
 = 2m s_+ \Delta q_{\rm con}
\end{equation}
and
\begin{equation}
\langle p,s|K_+(0)|p,s\rangle _c
 = 2m s_+ \biggl( - {\alpha_s \over 2 \pi} \Delta g
\biggr).
\end{equation}
(Here $\alpha_s = {g^2 \over 4 \pi}$.) 
To obtain the parton model description of $g_A^{(0)}$
the factor $- {\alpha_s \over 2 \pi}$ is then disassociated
with the operator $K_+$ and reassociated with the first moment
of the coefficient function of the polarised gluon distribution, 
viz. \cite{efremov,ccm}
\begin{equation}
\int_0^1 C^{(g)}_{\bf A_+=0} = - {\alpha_s \over 2 \pi}.
\end{equation}
The gluonic term in Eq.(27) 
offers a 
possible source for the OZI violation in $g_A^{(0)}|_{\rm inv}$.

In the parton model, the axial anomaly provides a local measurement
of the gluon polarisation $\Delta g$.
If we calculate the first moment of the box graph for photon-gluon
fusion then we find the sum of two contributions \cite{bint}:
\begin{equation}
\int_0^1 dx g_1^{\gamma^{*} g} =
- {\alpha_s \over 2 \pi}
\left[1-
\frac{2m_{q}^{2}}{p^{2}}
\frac{1}{\sqrt{1-4m_{q}^{2}/p^{2}}}
\ln \left(
\frac{1 - \sqrt{1-4m_{q}^{2}/p^{2}}}{1+\sqrt{1-4m_{q}^{2}/p^{2}}}
\right)\right].
\end{equation}
Here $-p^2$ is the virtuality of the gluon and $m_q$ 
is the mass 
of the quark which is liberated into the final state.
(The photon-gluon fusion is calculated in the Bjorken limit 
 $Q^2 \gg -p^2, m_q^2$.)
The unity term is the anomaly. It comes from the region of
phase space where the hard photon scatters on a quark or
antiquark
carrying transverse momentum squared $k_T^2 \sim Q^2$ \
\footnote{The transverse momentum is defined as orthogonal
 to the two-dimensional plane spanned by the momenta of
 the incident photon and the target gluon.}.
The second term comes from the kinematic region
$k_T^2 \sim -p^2,m_q^2$ 
--- it is the gluon matrix element of 
$m_q \left[ {\overline q} i \gamma_5 q \right]$.
If we make a cut-off $\lambda$ on the quark transverse momentum
where
$Q^2 \gg \lambda^2 \gg -p^2, m_q^2$, then we recover just 
the anomaly
term as a local measurement of the gluon's spin \cite{ccm}.
If we make a cut-off on, say, the quark virtuality $-k^2$ 
which mixes
transverse and longitudinal momentum components, then we obtain 
half of the anomaly in the corresponding gluon coefficient
\cite{bint}
when the cut-off $\lambda$ is chosen such that
$Q^2 \gg \lambda^2 \gg -p^2, m_q^2$.
The factorisation scheme dependence of 
the decomposition of $G_A(0)$ into quark and gluonic 
contributions has
been studied at length in Refs.\cite{bint,man90,bodwin}.
The transverse momentum cut-off approach provides the most
natural link to the anomaly.

If we were to work only in the light-cone gauge we might think
that we have a complete parton model 
description of the first moment of $g_1$.
However, one is free to work in any gauge including a covariant
gauge where the forward matrix elements of $K_+$ are not
invariant under ``large'' gauge transformations.
It remains an open question whether the net non-perturbative
quantity which
is shuffled between $K_A(0)$ and $(G_A - K_A)(0)$ under ``large''
gauge transformations
is finite or not.
If it is finite and, therefore, physical, then, when we choose
$A_+ =0$,
it must be frozen into some combination of $\Delta q_{\rm con}$
and $\Delta g$ in Eqs.(27-29).

The topological winding number depends on the gluonic boundary
conditions at infinity.
It is insensitive to local deformations of the gluon 
field $A_{\mu}(z)$ or of the gauge transformation $U(z)$.
When we take the Fourier transform to momentum space 
the topological structure induces a light-cone zero-mode which 
can contribute to $g_1$ only at $x=0$.
Hence, we are led
to consider the possibility that there may be a 
term in $g_1$  which is proportional to $\delta(x)$.

It is worthwhile to stop and contrast the two different infra-red
problems that one has to worry about when understanding the first
moment of $g_1$.
The factorisation scheme dependence of the decomposition of 
$G_A(0)$ into a quark term and a gluonic term is a problem in 
perturbative QCD, 
which is derived from the QCD Lagrangian by demanding invariance 
under ``small'' gauge transformations.
The problem of the (non-)invariance of the forward matrix elements
of $K_+$ is about the topological structure of the QCD vacuum 
--- that is, invariance under ``large'' gauge transformations.

\section{The polarised condensate and $U_A(1)$ symmetry}

We now explain how tunneling processes may induce a polarised
condensate inside a nucleon.  The formation of this polarised 
condensate is related to the realisation of $U_A(1)$ symmetry 
breaking \cite{rjc,thooft,thrept} by instantons.

\subsection{Realisations of $U_A(1)$ symmetry}

Let us choose $A_0=0$ gauge and define two operator charges:
\begin{equation}
X(t) = \int d^3z J_{05}^{GI}(z)
\end{equation}
and
\begin{equation}
Q_5(t) = \int d^3z J_{05}^{\rm con}(z)
\end{equation}
corresponding to the gauge-invariant 
and partially conserved axial-vector currents respectively.

When topological effects are taken into account, 
the QCD vacuum $|\theta \rangle$
is a coherent superposition 
\begin{equation}
|\theta \rangle = \sum_m {\rm e}^{im \theta} |m \rangle
\end{equation}
of the
eigenstates $|m \rangle$ of 
$\int d \sigma_{\mu} K^{\mu} \neq 0$ \cite{Callan,crewther}
(for a recent review, see \cite{shifman}).
Here
$\sigma_{\mu}$ is a large surface which is defined 
\cite{crewther} such that its boundary is spacelike 
with respect to the positions 
$z_k$ of
any operators or fields in the physical problem under discussion.
For integer values of the topological winding number $m$,
the states $|m \rangle$ contain $mf$ quark-antiquark 
pairs with non-zero $Q_5$ chirality
$\sum_l \chi_l = - 2 f m$ 
where $f$ is the number of light-quark flavours.
Relative to the $|m=0 \rangle$ state, the $|m=+1 \rangle$ 
state carries topological winding number +1 and $f$
quark-antiquark pairs with $Q_5$ chirality equal to $-2f$.

There are two schools of thought 
\cite{rjc,thrept}
about how instantons break $U_A(1)$ symmetry.
Both of these schools start from 't Hooft's 
observation \cite{thooft} 
that the flavour determinant
\begin{equation}
\langle {\rm det} \biggl[
{\rm {\overline q}_L}^i {\rm q_R}^j {\rm (z)} 
\biggr]
\rangle_{\rm inst.} \neq 0
\end{equation}
in the presence of a vacuum tunneling process 
between states with different topological winding number.
(We denote the tunneling process by the subscript ``inst.''.
 It is not specified at this stage whether ``inst.'' denotes 
 an instanton or an anti-instanton.)

\begin{enumerate}
\item
{\bf Explicit $U_A(1)$ symmetry breaking} 

In this scenario \cite{thooft,thrept} the $U_A(1)$ symmetry 
is associated with the current $J_{\mu 5}^{GI}$ and 
the topological charge density is treated like a mass term 
in the divergence of $J_{\mu 5}^{GI}$.
The quark chiralities which appear in the flavour 
determinant (35) are associated with $X(t)$
so that the net axial charge $g_A^{(0)}$ is not conserved 
$(\Delta X \neq 0)$
and the net $Q_5$ chirality is conserved $(\Delta Q_5 = 0)$ 
in quark instanton
scattering processes.

In QCD with $f$ light flavoured quarks the (anti-)instanton 
``vertex''  involves a total of $2f$ light quarks and 
antiquarks. 
Consider a flavour-singlet combination of $f$ right-handed 
($Q_5 =+1$) quarks incident on an anti-instanton.
The final state for this process consists of a
flavour-singlet combination of
$f$ left-handed ($Q_5 = -1$) quarks;
$+2f$ units of $Q_5$ chirality are taken away by an effective 
``schizon'' which carries zero energy and zero momentum 
\cite{thrept}.
The ``schizon'' is introduced to ensure $Q_5$ conservation.
The non-conservation of $g_A^{(0)}$ is ensured by a term coupled 
to $K_{\mu}$ with
equal magnitude and opposite sign to the ``schizon'' 
term which 
also carries zero energy and zero momentum.
This gluonic term describes the change in the topological
winding number which is induced by the tunneling process.
The anti-instanton changes the net $U_A(1)$ chirality by 
an amount $(\Delta X = -2f)$.

This picture is the basis of 't Hooft's effective instanton
interaction \cite{thooft}.

\item
{\bf Spontaneous $U_A(1)$ symmetry breaking} 

In this scenario \cite{rjc,crewther} the $U_A(1)$ symmetry is 
associated with the partially-conserved axial-vector current 
$J_{\mu 5}^{\rm con}$. 
Here, the quark chiralities which appear in the flavour 
determinant (35) are identified with $Q_5$. 
With this identification,
the net axial charge $g_A^{(0)}$ is conserved 
$(\Delta X = 0)$
and the
net $Q_5$ chirality is not conserved $(\Delta Q_5 \neq 0)$ 
in 
quark instanton
scattering processes.
This result is the opposite to what happens in the explicit
symmetry breaking scenario.
When $f$ right-handed quarks scatter on an instanton 
\footnote{cf. an anti-instanton in the explicit $U_A(1)$ 
 symmetry breaking scenario.}
the final state involves $f$ left-handed quarks.
There is no ``schizon'' and the instanton induces a change 
in the net $Q_5$ 
chirality
$\Delta Q_5 = -2f$.
The conservation of $g_A^{(0)}$ is ensured by the gluonic term 
coupled to $K_{\mu}$ 
which measures the change in the topological winding number
and which carries zero energy and zero momentum.
The charge $Q_5$ is time independent for massless quarks 
(where $J_{\mu 5}^{\rm con}$ is conserved).
Since $\Delta Q_5 \neq 0$ in quark instanton scattering 
processes we find that the $U_A(1)$ symmetry is spontaneously 
broken by instantons.
The Goldstone boson is manifest \cite{rjc} as the massless 
Kogut-Susskind pole which 
couples to $J_{\mu 5}^{\rm con}$ and $K_{\mu}$ but not to $J_{\mu 5}^{GI}$
--- see Eq.(22).

\end{enumerate}
In the rest of this paper we explain why these two 
possible realisations of $U_A(1)$ symmetry have a 
different signature in $\nu p$ elastic scattering.

\subsection{Formation of the polarised condensate}

In both the explicit and spontaneous symmetry breaking scenarios
we may consider multiple scattering of the incident quark first 
from an instanton and then from an anti-instanton.
Let this process recur a large number of times.
When we time-average over a large number of such
interactions, then the time averaged expectation
value of the chirality $Q_5$
carried by the incident quark is reduced from the naive value $+1$
that it would take in the absence of vacuum tunneling processes.
Indeed, in one flavour QCD the time averaged value of $Q_5$ tends 
to zero at large times \cite{forte2,forte1}.

In the spontaneous $U_A(1)$ symmetry breaking scenario \cite{rjc}
any instanton induced suppression of the flavour-singlet axial 
charge which is measured in polarised deep inelastic scattering 
is compensated by a net transfer of axial charge or ``spin'' from 
partons carrying finite momentum fraction $x$ to a polarised 
$U_A(1)$ condensate at $x=0$.
The polarised condensate induces a flavour-singlet 
$\delta (x)$ term in $g_1$ which is not present in 
the explicit $U_A(1)$ symmetry breaking scenario.
The net polarised condensate is gauge invariant.
In the $A_0=0$ gauge the condensate polarisation 
is ``gluonic'' 
and is measured by $\int d^3z K_0$.
In the light-cone gauge the polarisation of the condensate may 
be re-distributed between the ``quark'' and ``gluonic'' terms 
measured by $J_{+ 5}^{\rm con}$ and $K_+$ respectively.

\section{How to measure a polarised condensate}

Polarised deep inelastic scattering experiments measure $g_1$
between a small but finite value of $x$ and a value of $x$ 
which
is close to one. 
They cannot, even in principle, measure directly at $x=0$.
However, there are rigorous sum-rules for the first moment 
of $g_1$
for both proton and photon targets. 
These sum-rules include any contribution from the end-point $x=0$.

In the case of a polarised real photon target one finds \cite{photona}
\begin{equation}
\int_0^1 dx g_1^{\gamma} (x,Q^2) = 0.
\end{equation}
This sum-rule follows from electromagnetic gauge invariance
together with the absence of any massless pole
contributions to the photon
matrix
elements of the axial vector currents.
It has been generalised to a virtual photon target in \cite{photonb}.
If a polarised condensate exists in,
for example,
the hadronic (vector-meson dominance) part of the photon wavefunction,
then we
would expect to see a violation of this sum-rule in a polarised
deep
inelastic experiment.

For a proton target, 
the scale-invariant flavour singlet axial charge can be measured
independently in an elastic neutrino proton scattering experiment
\cite{garvey}.
Rigorous QCD renormalisation group arguments tell us that
the neutral current axial charge which is measured 
in elastic
$\nu p$ scattering is \cite{bcft}
\begin{equation}
g_A^{(Z)} =
{1 \over 2} g_A^{(3)} + {1 \over 6} g_A^{(8)} 
- {1 \over 6} (1 + {\cal C})
g_A^{(0)}|_{\rm inv}  + {\cal O}({1 \over m_h}).
\end{equation}
Here ${\cal C}$ denotes the leading order heavy-quark 
contributions 
to $g_A^{(Z)}$ and $m_h$ is the heavy-quark mass.
Numerically,
${\cal C}$ is a $\simeq 6-10\%$ correction \cite{bcft,kaplan}
---
within the present experimental error on $g_A^{(0)}|_{\rm inv}$.
The flavour-singlet axial charge in Eq.(37) includes
any contribution
from a polarised condensate
\footnote{
Heavy-quark condensates contribute only at ${\cal O}(1/m_h)$
in the heavy-quark mass $m_h$ \cite{svz}.
It follows that the coefficient of
any heavy-quark $\delta(x)$
term in $g_1$ decouples as ${\cal O}(1/m_h)$.
It does not affect our derivation of the relation between polarised
deep inelastic scattering and $\nu p$ elastic scattering.}
.

A polarised $U_A(1)$ condensate contributes to the value of
$g_A^{(0)}|_{\rm inv}$ which is extracted
from $\nu p$ elastic scattering
but not to the flavour-singlet axial charge which is extracted 
from  a polarised deep inelastic experiment.
A precise measurement of $\nu p$ elastic scattering would
make an important contribution to our understanding of
the internal spin structure of the nucleon and help to
resolve the issue of spontaneous versus explicit breaking of
$U_A(1)$ symmetry by instantons.

\vspace{1.0cm}
{\large \bf Acknowledgements: \\}
\vspace{3ex}

It is a pleasure to thank S.J. Brodsky, R.J. Crewther, T. Goldman, 
V.N. Gribov, P. Minkowski, E. Reya and A.W. Thomas 
for helpful discussions.
This work was supported by the Alexander von Humboldt Foundation.

\vspace{8.0cm}

\pagebreak

\end{document}